\documentclass[acmtog]{acmart}
\acmSubmissionID{-}

\usepackage{booktabs} 
\usepackage{multirow}
\usepackage{array}
\usepackage{tabularx}
\usepackage{comment}
\newcolumntype{L}[1]{>{\raggedright\let\newline\\\arraybackslash\hspace{0pt}}m{#1}}
\newcolumntype{C}[1]{>{\centering\let\newline\\\arraybackslash\hspace{0pt}}m{#1}}
\newcolumntype{R}[1]{>{\raggedleft\let\newline\\\arraybackslash\hspace{0pt}}m{#1}}

\usepackage[ruled]{algorithm2e} 

\SetAlFnt{\small}
\SetAlCapFnt{\small}
\SetAlCapNameFnt{\small}
\SetAlCapHSkip{0pt}

\acmJournal{TOG}
\acmVolume{41}
\acmNumber{6}
\acmArticle{222}
\acmYear{2022}
\acmMonth{12}

\setcopyright{rightsretained}

\acmDOI{10.1145/3550454.3555456}

\begin{document}
\title{Dressing Avatars: Deep Photorealistic Appearance for Physically Simulated Clothing}

\author{Donglai Xiang}
\affiliation{%
 \institution{Carnegie Mellon University}
 \country{USA}
}
\affiliation{%
 \institution{Meta Reality Labs Research}
 \country{USA}
}
\email{donglaix@cs.cmu.edu}

\author{Timur Bagautdinov}
\email{timurb@fb.com}
\author{Tuur Stuyck}
\email{tuur@fb.com}
\author{Fabian Prada}
\email{fabianprada@fb.com}
\affiliation{%
 \institution{Meta Reality Labs Research}
 \country{USA}
}

\author{Javier Romero}
\email{javierromero1@fb.com}
\author{Weipeng Xu}
\email{xuweipeng@fb.com}
\author{Shunsuke Saito}
\email{shunsukesaito@fb.com}
\affiliation{%
 \institution{Meta Reality Labs Research}
 \country{USA}
}
 
\author{Jingfan Guo}
\email{guo00109@umn.edu}
\affiliation{%
 \institution{University of Minnesota}
 \country{USA}
}

\author{Breannan Smith}
\email{breannan@fb.com}
\author{Takaaki Shiratori}
\email{tshiratori@fb.com}
\author{Yaser Sheikh}
\email{yasers@fb.com}
\affiliation{%
 \institution{Meta Reality Labs Research}
 \country{USA}
}
 
\author{Jessica Hodgins}
\affiliation{%
 \institution{Carnegie Mellon University}
 \country{USA}
}
\affiliation{%
 \institution{FAIR, Meta AI}
 \country{USA}
}
\email{jkh@cmu.edu}

\author{Chenglei Wu}
\email{chenglei@fb.com}
\affiliation{%
 \institution{Meta Reality Labs Research}
 \country{USA}
}

\renewcommand\shortauthors{Xiang, D. et al}

\begin{abstract}
Despite recent progress in developing animatable full-body avatars, realistic modeling of clothing - one of the core aspects of human self-expression - remains an open challenge. State-of-the-art physical simulation methods can generate realistically behaving clothing geometry at interactive rates. Modeling photorealistic appearance, however, usually requires physically-based rendering which is too expensive for interactive applications. On the other hand, data-driven deep appearance models are capable of efficiently producing realistic appearance, but struggle at synthesizing geometry of highly dynamic clothing and handling challenging body-clothing configurations. To this end, we introduce pose-driven avatars with explicit modeling of clothing that exhibit both photorealistic appearance learned from real-world data and realistic clothing dynamics. The key idea is to introduce a neural clothing appearance model that operates on top of explicit geometry: at training time we use high-fidelity tracking, whereas at animation time we rely on physically simulated geometry. Our core contribution is a physically-inspired appearance network, capable of generating photorealistic appearance with view-dependent and dynamic shadowing effects even for unseen body-clothing configurations. We conduct a thorough evaluation of our model and demonstrate diverse animation results on several subjects and different types of clothing. Unlike previous work on photorealistic full-body avatars, our approach can produce much richer dynamics and more realistic deformations even for many examples of loose clothing. We also demonstrate that our formulation naturally allows clothing to be used with avatars of different people while staying fully animatable, thus enabling, for the first time, photorealistic avatars with novel clothing.
\end{abstract}

%




\begin{teaserfigure}
\centering
\includegraphics[width=7.05in]{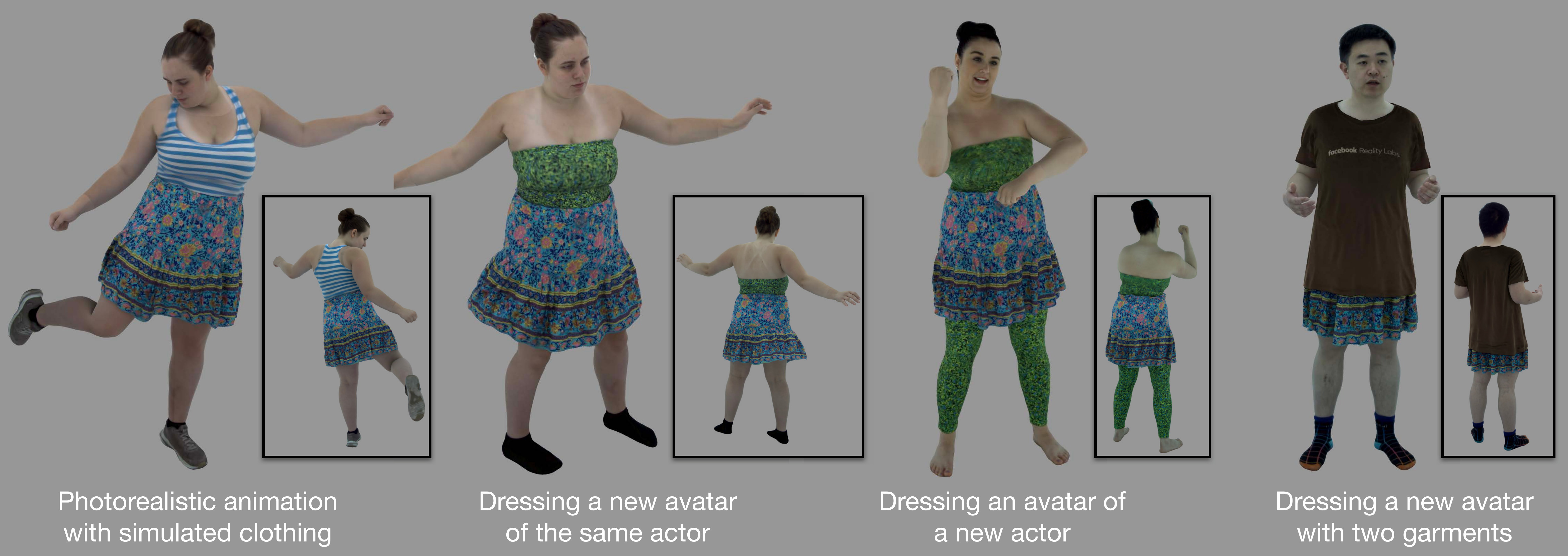}
\caption{We develop pose-driven full-body avatars with photorealistic clothing by applying neural rendering to physically simulated garments. On the left, we show a skirt animation together with the body avatar built from the same captured sequence. We further retarget the skirt to a novel sequence with the same actor and two new actors. On the right, we animate the skirt and a T-shirt together.}
\label{fig:teaser}
\end{teaserfigure}

\maketitle

\section{Introduction}

Photorealistic avatars are an essential component of authentic social telepresence, which is one of the key applications for AR/VR. Recent work on full-body avatars~\cite{bagautdinov2021driving,xiang2021modeling,habermann2021deeper} 
introduced animatable models capable of producing photorealistic synthetic representations of humans from sparse signals such as body pose.
%
%
However, synthesizing realistic clothing still remains a challenge in
avatar modeling. At the same time, clothing is an essential form of self-expression, creating a fundamental need for high-quality photorealistic animation of clothing.

Existing work on avatars with animatable clothing can be categorized into two main 
streams.
Cloth simulation creates realistic clothing deformations with 
dynamics~\cite{bridson2002robust,muller2007position,bouaziz2014projective,macklin2016xpbd,ly2020projective},
but only focuses on modeling geometry. The other line of the work leverages real-world captures to build neural representations of clothing geometry~\cite{bertiche2021pbns} and may include appearance~\cite{habermann2021deeper,xiang2021modeling,liu2021neural}.
However, these systems usually damp the clothing dynamics, struggle at generalizing to unseen poses and cannot handle collisions well. Our key insight is that these two lines of work are actually complementary to each other, and combining them can help achieve the best of both worlds.

In this work, we propose to integrate physics-based cloth simulation into avatar modeling, so that the clothing on the avatar can be animated photorealistically with the body, while achieving high-quality dynamics, collision handling and capabilities to animate and render avatars with novel clothing. Our work builds upon full-body Codec Avatars~\cite{bagautdinov2021driving, xiang2021modeling}, which leverage a Variational Autoencoder (VAE) to model the geometry and appearance of a human body. In particular, we follow the multi-layer formulation of~\cite{xiang2021modeling}, but redesign the clothing layer to integrate a physically-based simulator. Namely, at the training stage, we learn the clothing appearance model using real-world data, by processing raw captures with our dynamic clothing registration pipeline. At test time, we simulate the clothing geometry on top of the underlying body model with appropriate material parameters, and then apply the learned appearance model to synthesize the final output.


Unfortunately, there are two major issues with a naive implementation of this pipeline. 
First, there exists a gap between the simulator output and the tracking obtained from the real data. Estimating the full set of physical parameters for body and clothing to faithfully reproduce the clothed body configuration remains an unsolved problem, despite some progress in controlled settings~\cite{miguel2012est} or in estimating only the body parameters~\cite{guo2021inverse}. There are inevitable differences between the test-time simulation output with manually selected parameters and the real-world clothing geometry used for training. Second, tracking clothing and underlying body geometry at high accuracy is still a challenging problem, especially for loose clothing such as skirts and dresses. Both of these issues, inconsistency between training and test scenarios and unreliable tracking, make learning a generalizable appearance model 
more challenging. Thus, a good design of the appearance model should avoid learning chance 
correlations between degenerated tracked cloth geometry and specific appearance. To this end, we design the model to be localized in terms of both architecture (U-Net~\cite{unet_miccai2015}) and input representation (normals). We also take inspiration from physically-based rendering and decompose appearance into local diffuse components, view-dependent and global illumination effects such as shadowing. In particular, we rely on an unsupervised shadow network conditioned on the ambient occlusion map explicitly computed from the body and clothing geometry, so that the dynamic shadowing can be effectively modeled even for a different 
underlying body model at test time.    


Our approach generates physically realistic dynamics and photorealistic appearance that are robust to diverse body motion with complex body-clothing interactions. In addition, our formulation allows the transfer of clothing between different individuals' body avatars as shown in Fig.~\ref{fig:teaser}, as well as virtual garment resizing in Fig.~\ref{fig:edit-size}. Our method opens up the possibility to dress photorealistic avatars with novel garments. To summarize, our contributions are as follows:

\begin{itemize}
    \item We present animatable clothed human avatars with data-driven photorealistic appearance and physically realistic clothing dynamics from simulation;
    \item We develop a deep clothing appearance model to produce photorealistic clothing appearance that bridges the generalization gap between the tracked clothing geometry for training and the simulated clothing geometry at test time;
    \item Our animation system further enables transferring clothing between different subjects as well as editing of garment size.
\end{itemize}

In our experiments, we evaluate the effectiveness of our approach by animating multiple identities and clothing types, and provide a comprehensive qualitative and quantitative comparison to existing techniques.

\section{Related Work}

Our goal in this work is to build photorealistic animatable avatars, and we review work in this area in Sec. \ref{sec:rw_avatars}. To collect data for training such avatars we use techniques for multi-view clothing capture (Sec. \ref{sec:rw_capture}). We leverage physics-based clothing animation (Sec. \ref{sec:rw_animation}) so that the animation output not only looks photorealistic in appearance, but also natural in terms of physical dynamics.

\subsection{Photorealistic Animatable Avatars}
\label{sec:rw_avatars}

Models for dynamic full-body humans have been widely studied because of their applications in the movie and gaming industries, as well as virtual and augmented reality. Human avatars are usually built to be animatable with joint angles and skinning to allow easy control of skeleton-level deformation. The classical approach for modeling full-body appearance is to use a fixed texture associated with the template mesh \cite{carranza2003free,starck2007capture,de2008performance}, which enables the rendering of the avatars by various graphics techniques such as direct shading or ray tracing.
However, it is impossible for simple rendering to account for the complicated appearance change under the variation of viewpoint, deformation, material and lighting.
Improved realism has been achieved in classic methods with techniques related to image-based warping~\cite{einarsson2006flowed,eisemann2008floating,jain2010moviereshape,volino2014optimal}, but they do not enable body reposing.
To achieve photorealistic rendering of animatable full-body human appearance using classical techniques requires heavy computation and the work of digital artists.
In recent years there has been a lot of interest in building photorealistic human avatars directly from captured images in a data-driven manner. In these approaches, a deep neural network takes as input a specific body configuration (for example pose or mesh vertices), and outputs a predicted image supervised by the ground truth capture from the same viewpoint \cite{shysheya2019textured,prokudin2021smplpix,habermann2021real,zhi2020texmesh}. 
Data-driven approaches based on image translation networks have also attacked this problem purely from a monocular 2D perspective, using rendered skeletons~\cite{chan2019everybodydancenow} or 2D body correspondences~\cite{wang2018vid2vid} as input.
Some work further utilizes recent breakthroughs in neural rendering, such as Neural Radiance Fields \cite{mildenhall2020nerf,peng2021neural,peng2021animatable,su2021nerf,xu2021h,liu2021neural,noguchi2021narf,zheng2022structured,kwon2021neural} and Deferred Neural Rendering \cite{thies2019deferred,raj2021anr,grigorev2021stylepeople}. Notably, full-body Codec Avatars \cite{bagautdinov2021driving,xiang2021modeling,remelli2022drivable} have been proposed as a promising approach to achieving photorealistic telepresence. Most of these approaches consider clothing as rigidly attached to the human body, and the results are limited to tight clothing such as T-shirts and pants. One exception is the work by Habermann and colleagues \shortcite{habermann2021real}, which can model the deformation of loose clothing such as skirts and dresses in a coarse-to-fine manner, but the predicted clothing deformation is not physically realistic because the approach lacks explicit physics priors.

\subsection{Clothing Capture}
\label{sec:rw_capture}

Clothing capture has been explored as a source of geometry for garment modeling. Previous work on this topic can be divided into multi-view approaches and single-view approaches. Multi-view approaches use a multi-camera system or a 3D scanner to capture and reconstruct the shape of the clothing.
In order for the captured clothing shape to be ready for use in downstream tasks, much effort has been devoted to registration, i.e, the representation of the garment geometry by a consistent mesh topology in a temporal sequence \cite{bradley2008markerless,pons2017clothcap,bhatnagar2019multi,hong2021garment4d}. Some work \cite{zhang2017detailed,chen2021tightcap} also estimates the underlying body pose and shape from the clothed human reconstruction. In this paper, multi-view clothing capture provides data to train our appearance modeling network.

Recent progress in computer vision makes it possible to reconstruct clothing shape from a single-view input, which allows a much simpler capture setup.
One line of work treats the problem as a monocular 3D shape regression task with explicit \cite{zheng2019deephuman,alldieck2019tex2shape} or implicit \cite{saito2019pifu,saito2020pifuhd,li2020monocular,xiu2021icon,huang2020arch,burov2021dynamic} shape representation. Alternatively, optimization-based \cite{xu2018monoperfcap,habermann2019livecap,xiang2020monoclothcap} or learning-based approaches \cite{habermann2020deepcap,habermann2021deeper} are used to deform a personalized template shape to match the input image. Most of these methods only focus on estimating clothed body shape without photorealistic texture (except \cite{li2020monocular}). In addition, due to the fundamental difficulty of estimating depth from monocular RGB images, the output accuracy of these approaches is deficient for high-quality telepresence. Higher depth accuracy can be achieved in a monocular setup by using depth cameras~\cite{chen2015depth,yu2018doublefusion}, although the quality of the results still does not match that of multiview systems.

\begin{figure*}[t]
    \centering
    \includegraphics[width=\linewidth]{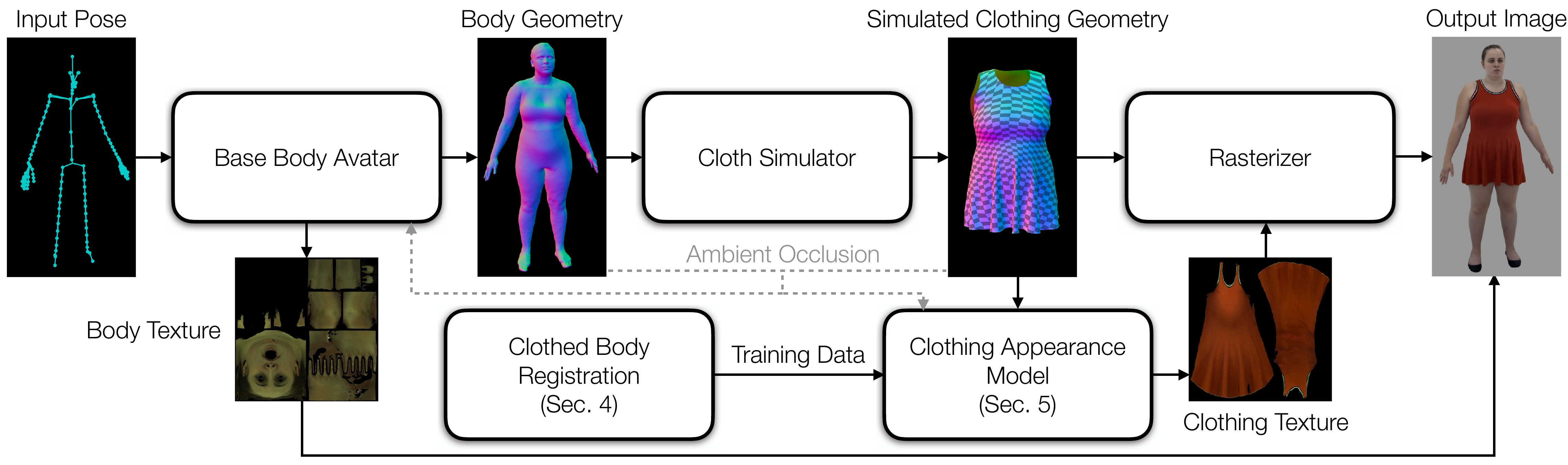}
    \caption{Our animation pipeline includes three major modules: the base body avatar that predicts body geometry and texture given pose as input, the cloth simulator that generates clothing deformation on top of the body geometry, and the clothing appearance model that predicts photorealistic clothing texture. The appearance model is trained using real captured data with registered body and clothing geometry. The body avatar and clothing appearance model also takes in ambient occlusion between the body and clothing geometry for dynamic shadowing effects. The geometry and texture pairs are then rasterized together to produce the final output image.}
    \label{fig:method_overview}
\end{figure*}

A notable line of work aims to capture clothing with the assistance of cloth simulation, both in the cases of monocular \cite{yang2017video,yang2018image,yu2019simulcap} and multiview input \cite{guo2021inverse,stoll2010animatable,li2021deep}. However, these methods are still limited in terms of output accuracy. Estimating precise physical parameters solely from visual input is still a difficult problem.

\subsection{Clothing Animation}
\label{sec:rw_animation}

Clothing is an integral part of human appearance, and thus the animation of clothing has long been studied as an important component in character animation. \textit{Physics-based simulation} of cloth has been established as a standard tool in the animation industry. Various aspects of cloth simulation, including speed, robustness, and collision detection and response, have been thoroughly studied \cite{baraff1998large,bridson2002robust,muller2007position,bouaziz2014projective,macklin2016xpbd, li2018implicit,ly2020projective,li2020codimensional}. In order to further improve animation efficiency, there have been efforts to combine physics-based animation with \textit{data-driven} approaches \cite{wang2010example,kim2013near}. A notable paradigm is to use simulation to provide training data to supervise machine learning models that predict clothing dynamics given body pose and shape inputs \cite{patel2020tailornet,santesteban2019learning,vidaurre2020fully,santesteban2021self,chentanez2020cloth,gundogdu2019garnet,zhang2021deep,bertiche2021deepsd,jin2020pixel,pfaff2021learnsimu,holden2019subspace}. The recent emergence of publicly available 3D captured clothing datasets \cite{ma2020learning,tiwari2020sizer} has inspired work that directly learns clothing animation models from such real-world data instead of simulation \cite{saito2021scanimate,ma2021scale,corona2021smplicit,ma2021power}. Additionally, the use of a self-supervised physics-based loss \cite{santesteban2022snug,bertiche2021pbns} has been explored for the generation of 3D garment deformations.

Although the aforementioned work can produce reasonable clothing shape deformation with dynamics, photorealistic appearance modeling of the animation output is a non-trivial problem especially for telepresence applications. Classic methods developed specialized BRDF models~\cite{yasuda92shading}, offline models for knitware like the lumislice~\cite{Xu2001lumislice}, or spatially-varying BRDFs which use pattern geometry arrays to improve realism like~\cite{daubert2001efficient}. More recently, classic graphic methods evolved to generate cloth fibers procedurally on the GPU in real-time, improving rendering quality~\cite{wu2019fiberlevel} although still not reaching photorealism. A more complete review of fabric rendering can be found in~\cite{castillo2019survey}. Among the learning-based methods, Xiang and colleagues \cite{xiang2021modeling} train a full-body avatar that models clothing as a separate layer for photorealistic appearance and temporal deformation, but they show only a single clothing style (T-shirt) and degree of dynamics. Zhang and colleagues \cite{zhang2021dynamic} propose a neural rendering approach that produces output images from coarse garments generated by a temporal model. However, this work has several limitations. First, this work shows results mostly in the synthetic domain. The output rendering is detailed but limited in photorealism, because generating highly photorealistic training images for the neural renderer is non-trivial. Second, the method assumes input pairs of body-only images as background and clothed body images as ground truth. In the real-world setting, it is not clear how to obtain such paired data, especially the body-only images. Third, the screen-space garment renderer can have difficulty inferring the depth order between the body in the background layer and the clothing in the foreground layer. By contrast, our method builds photorealistic avatars with highly dynamic clothing using captured data in the real world. We track body and clothing geometry and photometric correspondences at high quality in the captured data, which facilitates the effective learning of the appearance function. Our physically inspired clothing appearance model operates in 3D space, and thus is free from the depth ordering issue even with complex body-clothing interactions.

\begin{figure*}[t]
    \centering
    \includegraphics[width=\linewidth]{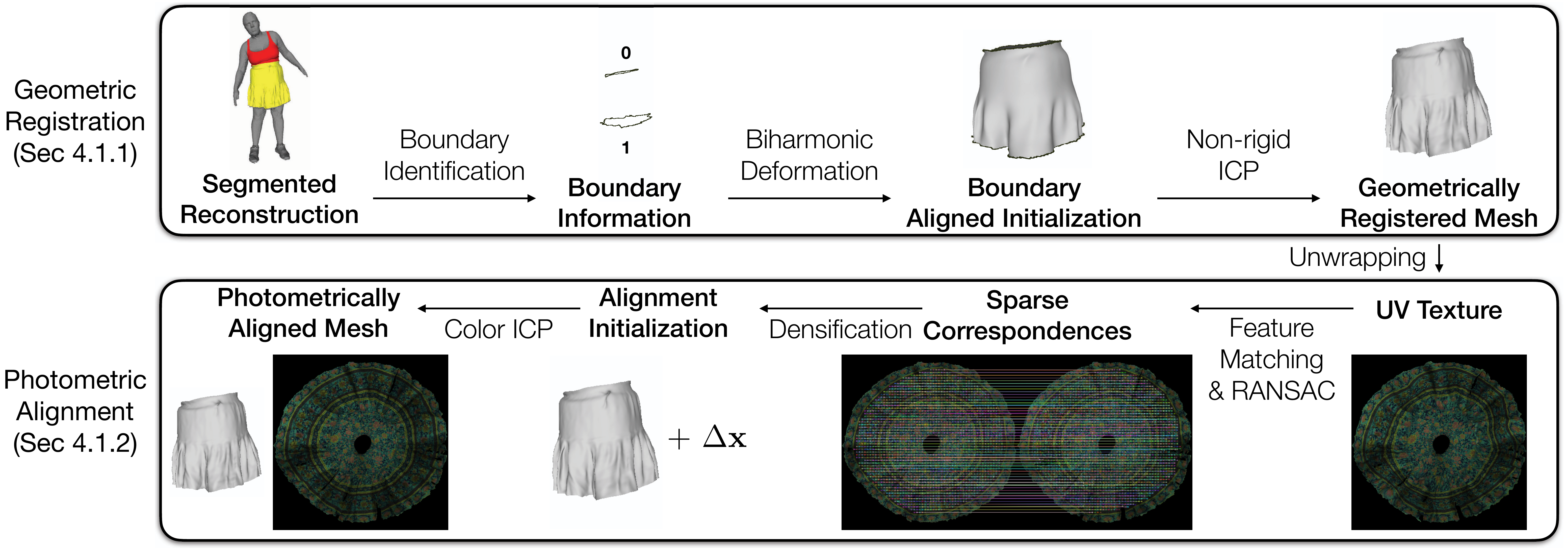}
    \caption{Our full clothing registration method consists of two major steps: geometric registration and photometric alignment. The geometric registration step aligns the garment surface of the segmented reconstruction with a template mesh. We rely on the boundary information for initialization and then minimize the surface error using non-rigid ICP. In the photometric alignment step, we establish sparse correspondences between the salient region of the unwrapped texture and a template texture. The sparse correspondences are then densified as a displacement for each vertex. This step is followed by color ICP which outputs both a geometrically and photometrically aligned mesh in the template topology.}
    \label{fig:method_tracking}
\end{figure*}

\section{Method Overview}

Our goal is to build pose-driven full-body avatars with dynamic clothing and
photorealistic appearance. Following~\cite{bagautdinov2021driving,xiang2021modeling}, we train our avatars on multi-view capture sequences of each subject wearing the clothing of interest. At test time, the avatars are animated by a sparse driving signal of skeleton motion (including facial keypoints if available), and can be rendered in a novel camera viewpoint.

We aim to achieve high-fidelity animation that looks realistic both in terms of appearance of the human subject and temporal clothing dynamics. For this purpose, we develop an animation pipeline consisting of three modules: an underlying body avatar model, physics-based cloth simulation and a clothing appearance model. The underlying body avatar takes as input the skeleton pose (including face conditioning if animating expression) and outputs the body geometry. Given a sequence of body geometry, we then use cloth simulation~\cite{stuyck2018cloth} to generate clothing geometry with natural and rich dynamics, physically 
consistent with the motion of the underlying body. Finally, we apply the clothing appearance model to the simulated geometry and generate photorealistic texture, which takes into account not only the clothing geometry but also shadows caused by the occlusion of the avatar body. The shadow cast by the clothing on the body is also modeled similarly in the underlying body avatar. Fig.~\ref{fig:method_overview} illustrates our overall pipeline.

Photorealistic full-body Codec Avatars~\cite{bagautdinov2021driving,xiang2021modeling} and physics-based cloth simulation~\cite{stuyck2018cloth} have been extensively explored in the existing literature. For these two modules we mostly follow previous work and provide implementation detail in the supplementary document\footnote{Unless otherwise stated, we use a GPU-based XPBD \cite{macklin2016xpbd} cloth simulation of a mass-spring system for its superior runtime performance, but our method is not restricted to a specific choice of model or integration technique. Examples generated by a different simulator \cite{ly2020projective} are provided in the supplementary document.}. We find efficient photorealistic appearance modeling of clothing to be a critical missing component in such a pipeline, and we tackle two major technical challenges in order to build such a system. On the one hand, as explained in Sec. \ref{sec:method_appearance_model}, we develop a deep clothing appearance model that can efficiently generate highly photorealistic clothing texture with dynamic view-dependent and shadowing effects. For the design of the model, we focus on the generalization of the produced appearance, because the input geometry from the cloth simulator can be different from the tracked clothing geometry used to train the appearance model. On the other hand, in order to generate training data for the appearance model, we extend the previous clothing registration algorithm \cite{xiang2021modeling,pons2017clothcap} to handle highly dynamic clothing types including a skirt and a dress. In addition, we track photometric correspondences on the garments by matching salient features, which are essential for the modeling of highly textured clothing appearance. The clothing registration step is described in Sec. \ref{sec:method_registration}.

\section{Clothed Body Registration}
\label{sec:method_registration}

In this section, we introduce our data processing pipeline to obtain training data for the clothing appearance model described in Sec.~\ref{sec:method_appearance_model}. Because clothing registration is not the core contribution of this paper, we focus on challenges posed by large clothing dynamics and rich texture, which must be addressed to achieve high-fidelity animation. 

Our data capture setup is similar to previous work~\cite{xiang2021modeling}. The pipeline takes multi-view image sequences of a subject as input, and outputs registered meshes of the garment and the underlying body in two separate layers. We follow \cite{xiang2021modeling} for raw geometry reconstruction and mesh segmentation into body and clothing regions. We similarly estimate the kinematic body pose and inner-layer body surface.

\begin{figure*}[t]
    \centering
    \includegraphics[width=\linewidth]{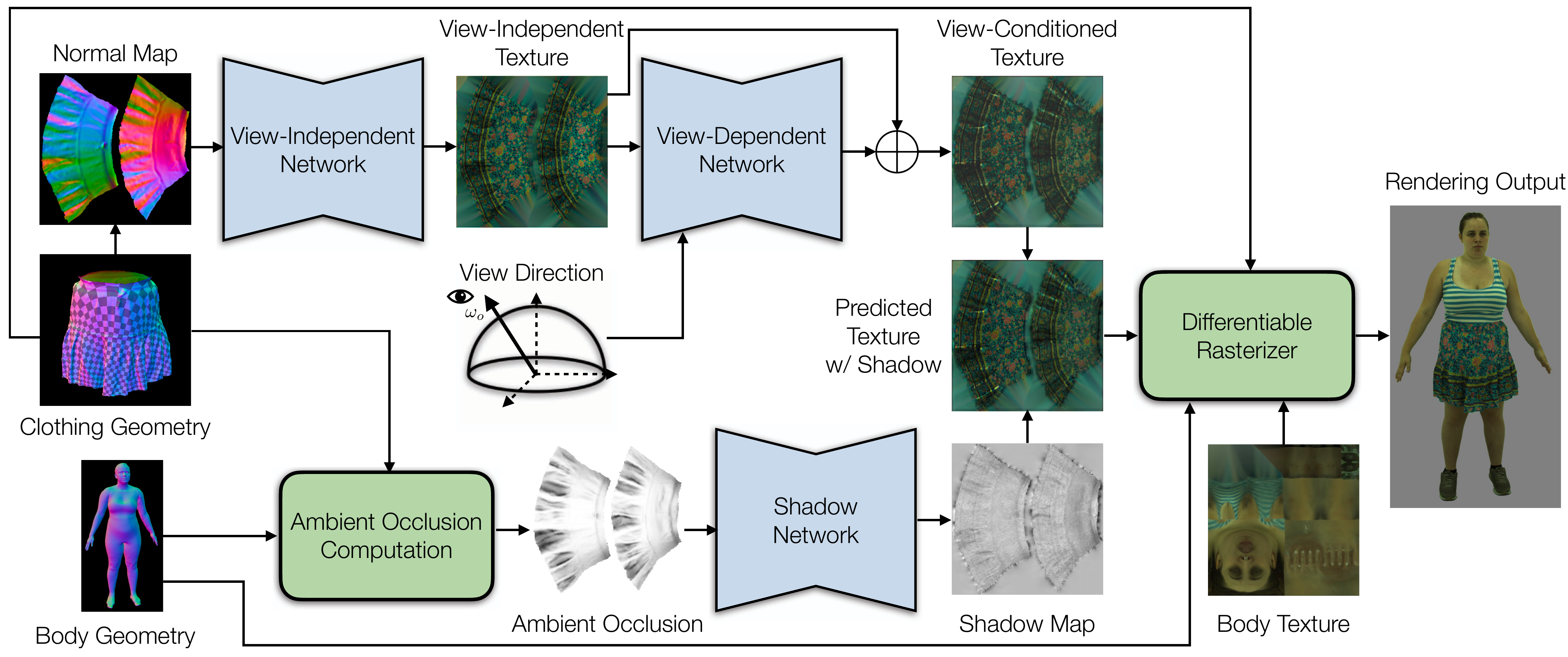}
    \caption{The architecture of our deep clothing appearance model. We model local diffuse appearance with a view-independent network conditioned on surface normals. The view-dependent network additionally takes in view direction and produces a view-conditioned offset. We explicitly model shadowing using a shadow network that predicts a multiplicative shadow map given ambient occlusion between the body and clothing as input.}
    \label{fig:appearance_model_architecture}
\end{figure*}

\subsection{Dynamic Clothing Registration}

The goal of clothing registration is to represent the clothing geometry in a single mesh topology with consistent correspondences. Our clothing registration method consists of two major steps, geometric registration and photometric alignment. Fig.~\ref{fig:method_tracking} illustrates the overview of the clothing registration method.

\subsubsection{Geometric Registration} In this step, we fit a clothing template to the segmented clothing region of the reconstructed mesh using the non-rigid Iterative Closest Point (ICP) algorithm. The non-rigid ICP algorithm is similar to previous work \cite{xiang2021modeling,pons2017clothcap} and thus we omit the details here.

In order to track loose and dynamic clothing types (e.g. skirt and dress), it is important to provide good initialization for non-rigid ICP. We observe that the clothing boundaries provide useful information about the overall orientation and deformation of the garment. Therefore, we start by estimating coarse boundary correspondences. Utilizing the fact that each garment has a fixed number of boundaries (for example, two for the skirt and four for the dress), we associate each point on the the target clothing boundary with the template mesh boundary by querying the nearest vertex in the tracked inner-body surface. Given this coarse boundary correspondence, we use Biharmonic Deformation Fields \cite{jacobson2010mixed} to solve for per-vertex deformations that satisfy the boundary alignment constraints while minimizing the interior distortion. We use the output of the Biharmonic Deformation Fields as the initialization for the non-rigid ICP algorithm.

\subsubsection{Photometric Alignment} The geometric registration method in \cite{xiang2021modeling,pons2017clothcap} aligns the clothing geometry with a single template topology by minimizing the surface distance, but does not explicitly solve for interior correspondences. In order to effectively model clothing appearance, it is necessary to make sure that each vertex in the template consistently tracks the same color (essentially reflectance), which we call photometric correspondences in this paper. We observe that the chunk-based inverse rendering algorithm in \cite{xiang2021modeling} can correct small deviations of the photometric correspondences in the geometric registration step but cannot recover from large errors in the initialization.

Here, we explicitly solve for photometric correspondences by matching salient features in highly textured regions of the garments. For each frame in the sequence, we first unwrap a mean texture from multi-view images to the UV space using the geometrically aligned mesh from the previous step. Then, following \cite{bogo2017dynamic}, we use DeepMatching \cite{revaud2016deepmatching} to establish sparse correspondence pairs between the unwrapped texture and the template texture. We also use RANSAC to prune erroneous correspondences. The sparse correspondences are then densified to each vertex by solving a Laplace's equation, similar to Biharmonic Deformation Fields in the previous step. Finally, we run color ICP (\cite{park2017colored}, our own implementation) between the template and the target mesh to photometrically align all the vertices.

\section{Deep Dynamic Clothing Appearance Model}
\label{sec:method_appearance_model}

In this section, we introduce our clothing appearance model which is the key technical component that enables our photorealistic clothing animation system. Given clothing geometry as input, the goal of the clothing appearance model is to generate clothing texture that can be used for rasterization together with the input geometry to produce photorealistic appearance.

We build a data-driven clothing appearance model in order to learn complex photorealistic appearance from real captured image sequences. Several factors need to be taken into account when designing such a model. First, the appearance model is trained on tracked geometry from the previous registration step, but at test time, it takes simulated clothing geometry as input. Therefore, it is essential for the model to bridge the generalization gap between the training and testing data. Second, the generated texture should include various aspects of photorealistic appearance, such as view-dependent effects and dynamic shadowing. Third, for the sake of efficiency, the model should only involve basic quantities that can be easily derived from the clothing geometry, without computationally expensive operations such as multi-bounce Monte-Carlo ray tracing.

\subsection{Background: Rendering Equation}

\label{sec:rendering_eq}

For a particular point $\mathbf x$ in the scene, the classical rendering equation \cite{kajiya1986rendering} is written as
\begin{gather}
    L_o(\mathbf x, \omega_o) = \int_{\Omega(\mathbf n)} f(\mathbf x, \omega_i, \omega_o) L_i(\mathbf x, \omega_i) (\omega_i \cdot \mathbf n) d\omega_i,
\end{gather}
where $\omega_i,\omega_o$ denotes the incident and outgoing direction, $L_i,L_o$ denotes the incident and output radiance, $f$ denotes the BRDF function, and $\Omega$ denotes the hemisphere where the integral is computed, determined by the normal direction $\mathbf n$. Here the emission term is omitted because clothing usually does not emit radiance.

A common strategy in physics-based rendering is to decompose the appearance into a diffuse component and a specular component. For the diffuse component, the BRDF function $f$ is assumed to be a constant per point. Then the diffuse radiance can be written as
\begin{gather}
    L_d(\mathbf x) = f_d(\mathbf x) \int_{\Omega(\mathbf n)} L_i(\omega_i) (\omega_i \cdot \mathbf n) d\omega_i.
\end{gather}
When assuming distant lighting and ignoring cast shadows, $L_i$ can be considered as an environment lighting map which represents the consistent illumination condition for the whole captured sequence. Under this circumstance, the integral term is only determined by the normal direction $\mathbf n$ at each point, and thus
\begin{gather}
    \label{eq:reflectance-shading}
    L_d(\mathbf x) = f_d(\mathbf x) E(\mathbf n), \quad E(\mathbf n) =  \int_{\Omega(\mathbf n)} L_i(\omega_i) (\omega_i \cdot \mathbf n) d\omega_i.
\end{gather}

Besides the local diffuse component, we need to additionally account for specular (view-dependent) effects and global illumination effects especially cast shadows. The decomposition of appearance into these three components has direct implication for the design of network architecture described in the following section.

\subsection{Model Formulation}

\label{sec:net_arch}

From our capture setup we obtain a sequence of registered clothing meshes $\{V_i\}$ and underlying body meshes $\{B_i\}$, paired with multi-view images ${I_i^c}$, where $i$ denotes the frame number and $c$ denotes the camera ID. We learn an appearance function $F$ from geometry $V_i, B_i$ and viewpoint $v^c$ to a photorealistic UV texture.
Fig.~\ref{fig:appearance_model_architecture} shows our clothing appearance model.

From the analysis in the previous section, the function $F$ should be able to model local diffuse (view-independent) appearance, view-dependent effects and cast shadows. Under the assumption in Eq.~\ref{eq:reflectance-shading}, the diffuse component at each location is determined by the normal direction and diffuse albedo. Therefore, the view-independent part of $F$ is conditioned on the normal direction $\mathbf n$ to encode the illumination direction.
We use 2D convolution with untied biases to encode the spatially varying reflectance over the clothing.
In our experiments, we show that the normal direction $\mathbf n$, as a local geometric property, is more effective for generalization when used as input conditioning than the absolute vertex positions $V_i$.

The view-dependent part of the network additionally takes the viewpoint information $v_c$ as input, and predicts an additive view-dependent appearance offset. The viewpoint is represented by the viewing direction vector in the local Tangent-Binormal-Normal (TBN) coordinate at each position on the mesh.

For dynamic shadowing effects, previous work \cite{xiang2021modeling,bagautdinov2021driving} demonstrates that shadow maps predicted by a shadow network from ambient occlusion can achieve a good tradeoff between output quality and computation efficiency. Therefore, we follow previous work to use a shadow network to predict a quasi-shadow map, which is multiplied with the view-dependent texture to obtain the final predicted texture. The input ambient occlusion is computed by ray-mesh intersection with both body and clothing geometry.

We use a 2D convolutional neural network (CNN) for the architecture of the appearance model $F$. All the inputs to the network are converted to 2D feature maps according to a fixed UV mapping. Instead of the autoencoder structure of Codec Avatars \cite{xiang2021modeling,bagautdinov2021driving}, we adopt the U-Net architecture~\cite{unet_miccai2015} in each of the three parts. This architecture can focus on local information around each input location to learn a generalizable appearance function.

\begin{table*}[t]
\centering
\caption{Quantitative evaluation of the clothing appearance model applied on tracked clothing geometry. Mean Squared Error (MSE, the lower the better) and Structural Similarity Index Measure (SSIM, the higher the better) are reported. We conduct ablation studies on the effectiveness of each network component, type of input conditioning, and photometric alignment for the training data. We also compare with similar modules in the previous work.}
\begin{tabular}{ L{11.5cm} | C{2.5cm} | C{2.5cm} }
    \hline 
    Method & MSE$\downarrow$ & SSIM$\uparrow$ \\
    \hline
    Network components: mean texture & 316.48 & 0.67 \\
    Network components: mean texture + shadow & 260.44 & 0.68 \\
    Network components: view-independent + shadow & 189.29 & 0.73 \\
    Network components: view-independent + view-dependent & 177.27 & 0.75 \\
    Our full model (network components: view-independent + view-dependent + shadow) & \textbf{155.31} & \textbf{0.76} \\
    \hline
    Geometry input conditioning: raw vertices & 206.10 & 0.72 \\
    Geometry input conditioning: unposed normal & 160.38 & \textbf{0.76} \\
    Geometry input conditioning: unposed vertices & 161.08 & \textbf{0.76} \\
    Our full model (geometry input conditioning: normal) & \textbf{155.31} & \textbf{0.76} \\
    \hline
    W/o photometric alignment for training data (test on data w/o photometric alignment) & 597.71 & 0.44 \\
    W/o photometric alignment for training data (test on data w/ photometric alignment) & 483.73 & 0.46 \\
    Our full model (w/ photometric alignment for training data) & \textbf{155.31} & \textbf{0.76} \\
    \hline
    Previous method: Dynamic Neural Garments \cite{zhang2021dynamic} & 326.58 & 0.57 \\
    Previous method: Clothing Codec Avatars \cite{xiang2021modeling} (texture only) & 261.28 & 0.64 \\
    Previous method: Clothing Codec Avatars \cite{xiang2021modeling} (geometry + texture) & 379.87 & 0.58 \\
    Our full model & \textbf{155.31} & \textbf{0.76} \\
    \hline
\end{tabular}
\label{table:quantitative_appearance}
\end{table*}

\textit{Loss function.} The appearance model produces a view-conditioned photorealistic texture for the clothing given the input clothing and underlying body geometry (used for ambient occlusion computation), denoted by $F(V_i, B_i, v_c)$. A differentiable rasterizer $R$ is then used to render body and clothing, denoted by $R(V_i, F(V_i, B_i, v_c))$. We penalize the difference between the output rendering and the raw captured image in the clothing region with the following differentiable rendering loss
\begin{gather}
l = \sum_{i,c} \Vert (R(V_i, F(V_i, B_i, v_c)) - I_i^c) \odot M_i^c \Vert_1,
\label{eq:app_loss}
\end{gather}
where $M_i^c$ denotes the mask of clothing region in $I_i^c$ obtained from image segmentation, and $\odot$ denotes element-wise multiplication. We also rasterize the underlying tracked body geometry with a mean unwrapped body texture together with the clothing for correct body-clothing occlusion, which is omitted in Eq.~\ref{eq:app_loss} for simplicity.

\begin{figure}[t]
    \centering
    \includegraphics[width=\linewidth]{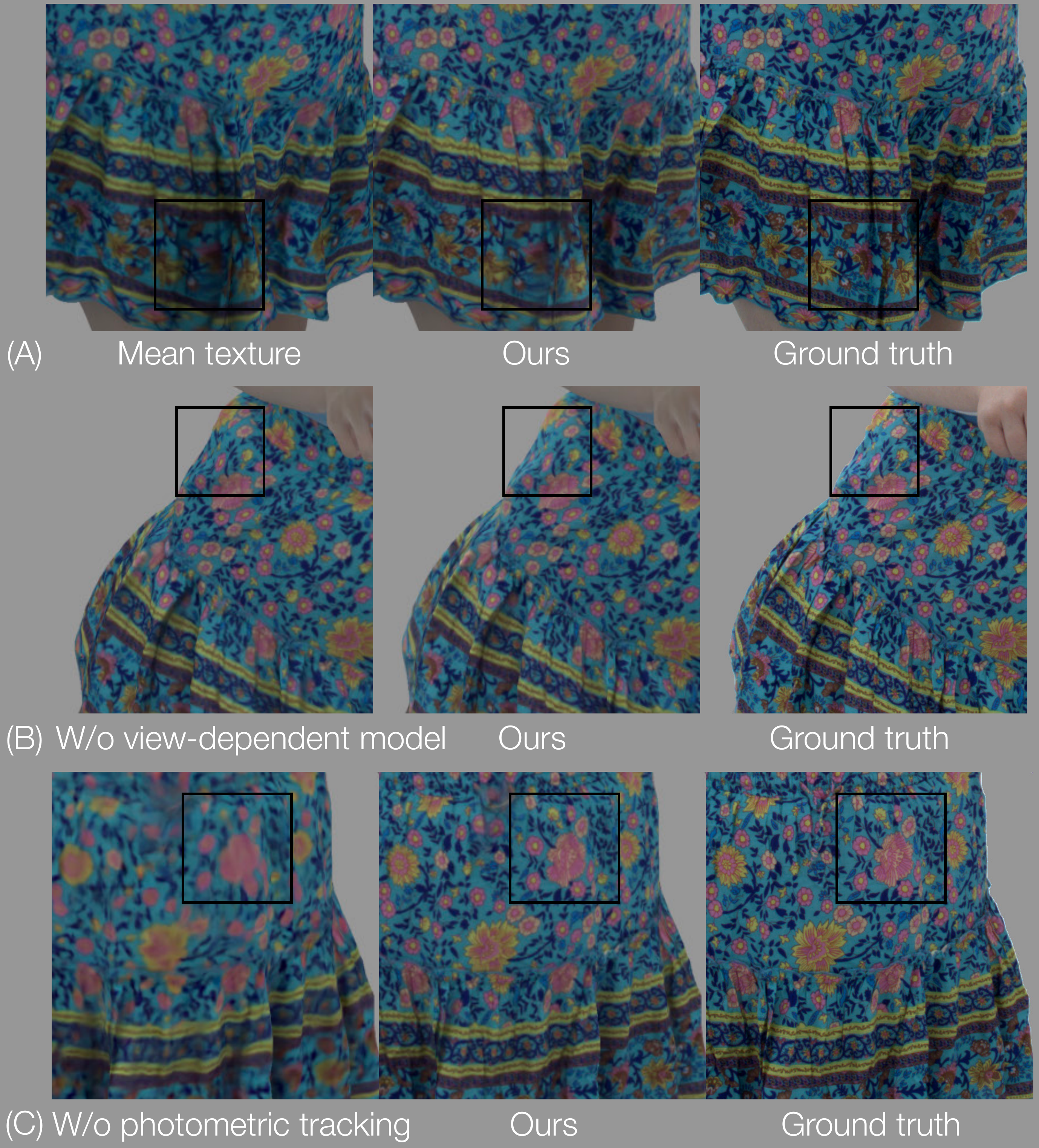}
    \caption{Illustration for ablation studies of the deep clothing appearance model. We compare our full model with representative baselines. Regions of particular interest are marked with black squares. (A) The mean-texture rendering has baked-in shadows. (B) Without the view-dependent component, brightness near the sihouette caused by Frensnel reflection cannot be modeled. (C) The model trained on data without photometric alignment produces blurry results.}
    \label{fig:ablation}
\end{figure}

\begin{figure}[t]
    \centering
    \includegraphics[width=\linewidth]{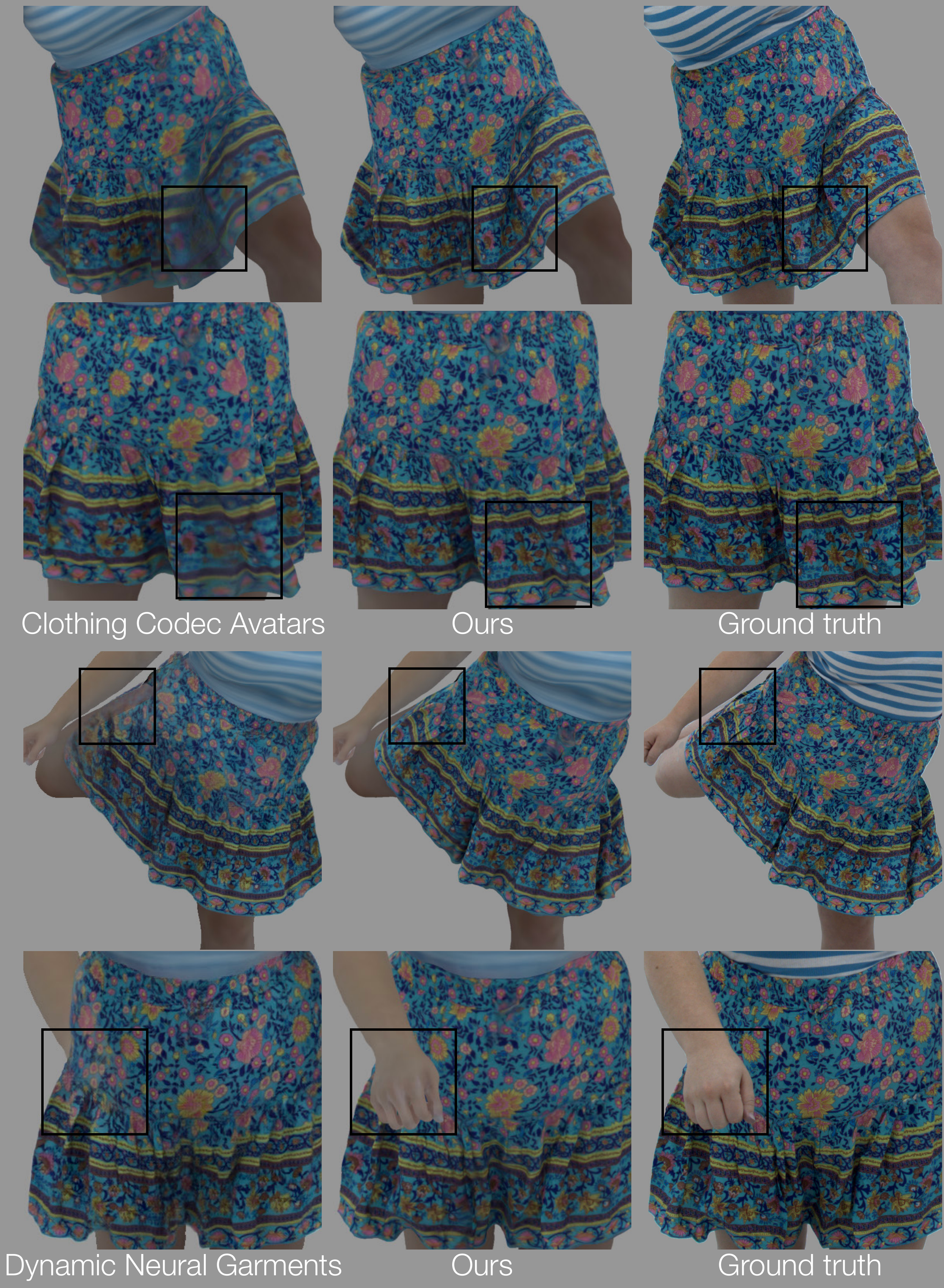}
    \caption{Comparison between our clothing appearance model and similar modules in previous methods. The Clothing Codec Avatars \cite{xiang2021modeling} have difficulties modeling the highly dynamic region of the skirt. A major issue with the screen-space garment renderer in Dynamic Neural Garments \cite{zhang2021dynamic} is the incorrect depth ordering between body in the background layer and clothing in the foreground layer.}
    \label{fig:app_prev}
\end{figure}

\section{Results}

In this section we present experimental results. We first introduce the data capture setup in Sec. \ref{sec:exp_setup}. Then we evaluate the clothing appearance model presented in Sec. \ref{sec:exp_app}, followed by full pose-driven animation results including body and clothing in Sec. \ref{sec:exp_animation}. In Sec. \ref{sec:exp_dressing}, we show an application of dressing photorealistic avatars, i.e. the animation of clothing on top of novel subjects. Finally, in Sec. \ref{sec:exp_runtime}, we present a runtime analysis of our method.

\subsection{Data Capture Setup}
\label{sec:exp_setup}

We capture and process a total of four garments: two T-shirts worn by two different male subjects, and a skirt and a dress worn by a female subject. The four garments span different types of texture patterns, including uniform color, a logo, and tiled floral texture, and thus provide a good testbed for our deep clothing appearance model. Following previous work \cite{xiang2021modeling}, we treat other clothing worn along with the four garments of interest in the body layer, including pants paired with the T-shirts and the tank top paired with the skirt. These garments move closely together with the body, and can be handled well in the same layer.

For the purpose of dressing avatars, we additionally capture three base body avatars with minimal clothing (tight outfits with greenish color in Fig.~\ref{fig:dressing}), including the subject from the skirt and dress capture and two new subjects. In Sec.~\ref{sec:exp_dressing}, we generate synthetic clothing animation on these three base body avatars.

\subsection{Evaluation of the Clothing Appearance Model}
\label{sec:exp_app}

In this section, we evaluate the deep clothing appearance model, including ablation studies and comparisons with previous work. We leave out a segment of the captured skirt sequence as the test set, and apply the clothing appearance model on the tracked clothing geometry. In this way, we can directly compute the difference between the rendered clothing output and the original captured images as ground truth. We report the error in the clothing area of the images indicated by the segmentation masks.

\subsubsection{Ablation studies: network components}

In the first group of experiments in Table~\ref{table:quantitative_appearance}, we validate the effectiveness of different modules in our clothing appearance model. The simplest version of the appearance model is to directly rasterize a single mean unwrapped texture for all the frames. This extremely cheap rendering method can retain the pattern on the skirt, but no dynamic appearance effects from illumination and shadowing are modeled, which is exemplified by the unnatural baked-in shadows in Row (A) of Fig.~\ref{fig:ablation}. We further add view-independent, view-dependent and shadow networks, which correspond to the three dynamic effects we attempt to model. In Row (B) of Fig.~\ref{fig:ablation}, we compare the full method with the network without view-dependent effects, which corresponds to the ``view-independent + shadow'' case in Table~\ref{table:quantitative_appearance}. The most obvious difference is the brightness pixel intensity near the silhouette of the clothing in the full method, due to Fresnel reflection when the incident angle is close to $90^{\circ}$. The results in Table~\ref{table:quantitative_appearance} verify that the full model with all three modules performs the best. 

\subsubsection{Ablation studies: input conditioning}
In Sec.~\ref{sec:net_arch}, our appearance model is conditioned on surface normals as geometry input. In the second group of experiments in Table~\ref{table:quantitative_appearance}, we compare this formulation with the one from \cite{xiang2021modeling}, where the network takes the clothing vertex location after inverse Linear Blend Skinning (LBS) as input, also called ``unposed vertices''. We additionally consider the other possibilities of using ``raw vertices'' (without inverse LBS) and the surface normals of unposed vertices (``unposed normal''). The quantitative results demonstrate the superiority of our proposed formulation. We believe there are two reasons for this performance improvement. First, the surface normals are a localized representation, and thus allow better generalization of the model than the absolute positions of vertices. Second, given the analysis in Sec.~\ref{sec:rendering_eq}, we know that the surface normals encode the direction of incident illumination, thus following the principles of physics-based rendering.

\subsubsection{Ablation studies: photometric alignment}
In the third group of experiments, we verify the importance of photometric alignment for highly textured clothing. As a comparison, we also train an appearance model using clothing meshes generated by geometric registration without photometric alignment. Both the quantitative results in Table~\ref{table:quantitative_appearance} and the qualitative evaluation in Row (C) of Fig.~\ref{fig:ablation} show significant degradation when photometric alignment is removed. The output images are very blurred when the model is trained on such data. Different positions in the textured regions of the clothing have a different reflectance (BRDF), so the tracked photometric correspondences must be consistently aligned with the captured images in order to learn the correct appearance function.

\subsubsection{Comparison with previous methods} We compare the clothing appearance model with similar modules from two previous methods: Clothing Codec Avatars \cite{xiang2021modeling} and Dynamic Neural Garments \cite{zhang2021dynamic}. For these experiments, we use the same training data as our full model and only test the influence of model choice.

Clothing Codec Avatars \cite{xiang2021modeling} are deep neural networks with an autoencoder architecture. The encoder takes unposed clothing geometry as input and outputs a latent code which is fed to the decoder to generate both geometry and photorealistic view-dependent texture, so the network can be used as clothing appearance model as well\footnote{In the original work, the encoder takes both geometry and texture as input. We adapt the encoder input for this experiment. The Clothing Codec Avatars have the same shadow network that takes ambient occlusion as input.}. We test the model in two ways: (1) rendering tracked geometry (identical to input) with decoded texture (2) rendering decoded geometry with decoded texture. As shown in the last group of results in \cite{xiang2021modeling} and in Fig.~\ref{fig:app_prev}, our deep appearance model achieves the lowest rendering errors, which we attribute to three factors. First, the bottleneck structure of Clothing Codec Avatars is essential for controlling the avatar with a low-dimensional latent code but disadvantageous for learning detailed appearance compared with the U-Net architecture of our appearance model. This analysis is further supported by comparing Clothing Codec Avatars with the appearance model with the same input conditioning (``unposed vertices'') but a U-Net architecture among the second group of experiments in Table~\ref{table:quantitative_appearance}. Second, Clothing Codec Avatars are trained using a differentiable rendering loss on decoded geometry and texture. However, the high frequency dynamics of the skirt itself is hard for the network to learn, further jeopardizing the modeling of texture, manifested by the blurry regions at the bottom of the skirt in Fig.~\ref{fig:app_prev}. Third, in the ablation studies, we verified that the input conditioning of unposed vertices is not as good as the surface normal used by our appearance model. All these factors demonstrate that for our goal, it is advantageous to adopt the novel formulation of the appearance model instead of that of Clothing Codec Avatars.

Dynamic Neural Garments (DNG) \cite{zhang2021dynamic} models garment appearance with an improved version of Deferred Neural Rendering \cite{thies2019deferred} that is more temporally coherent. The code for the garment renderer is released (not including the garment geometry prediction part), so we train the renderer using the same data as our appearance model for comparison. We generate the background images by rasterizing tracked body geometry with the fixed mean body texture, and composite it with the clothing region of the original image as ground truth. The results are shown in Table~\ref{table:quantitative_appearance} and Fig.~\ref{fig:app_prev}. As a screen-space method, the garment renderer in DNG has difficulty reasoning about the depth order between the body and clothing layer when there are complex interactions between the arms or legs and the garment. This limitation is acknowledged in the original paper \cite{zhang2021dynamic} and contributes to the high quantitative error in Table~\ref{table:quantitative_appearance} when skin color appears in the clothing region of ground truth images.

\begin{figure*}[t]
    \centering
    \includegraphics[width=\linewidth]{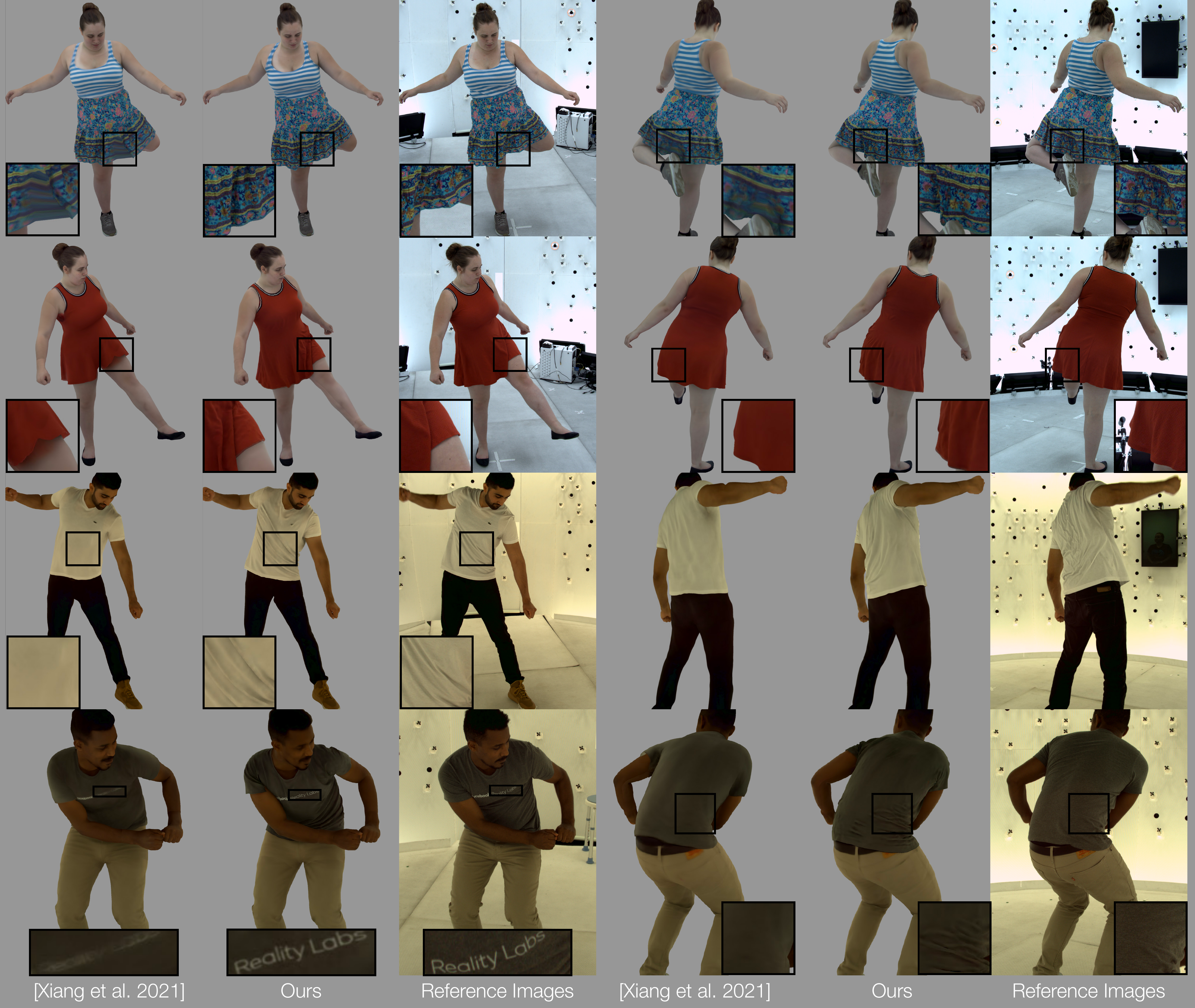}
    \caption{Pose-driven animation results in comparison with with \cite{xiang2021modeling}. In each row we show one result generated by both methods and a held out captured image for reference from two different views.}
    \label{fig:animation}
\end{figure*}

\begin{figure*}[t]
    \centering
    \includegraphics[width=\linewidth]{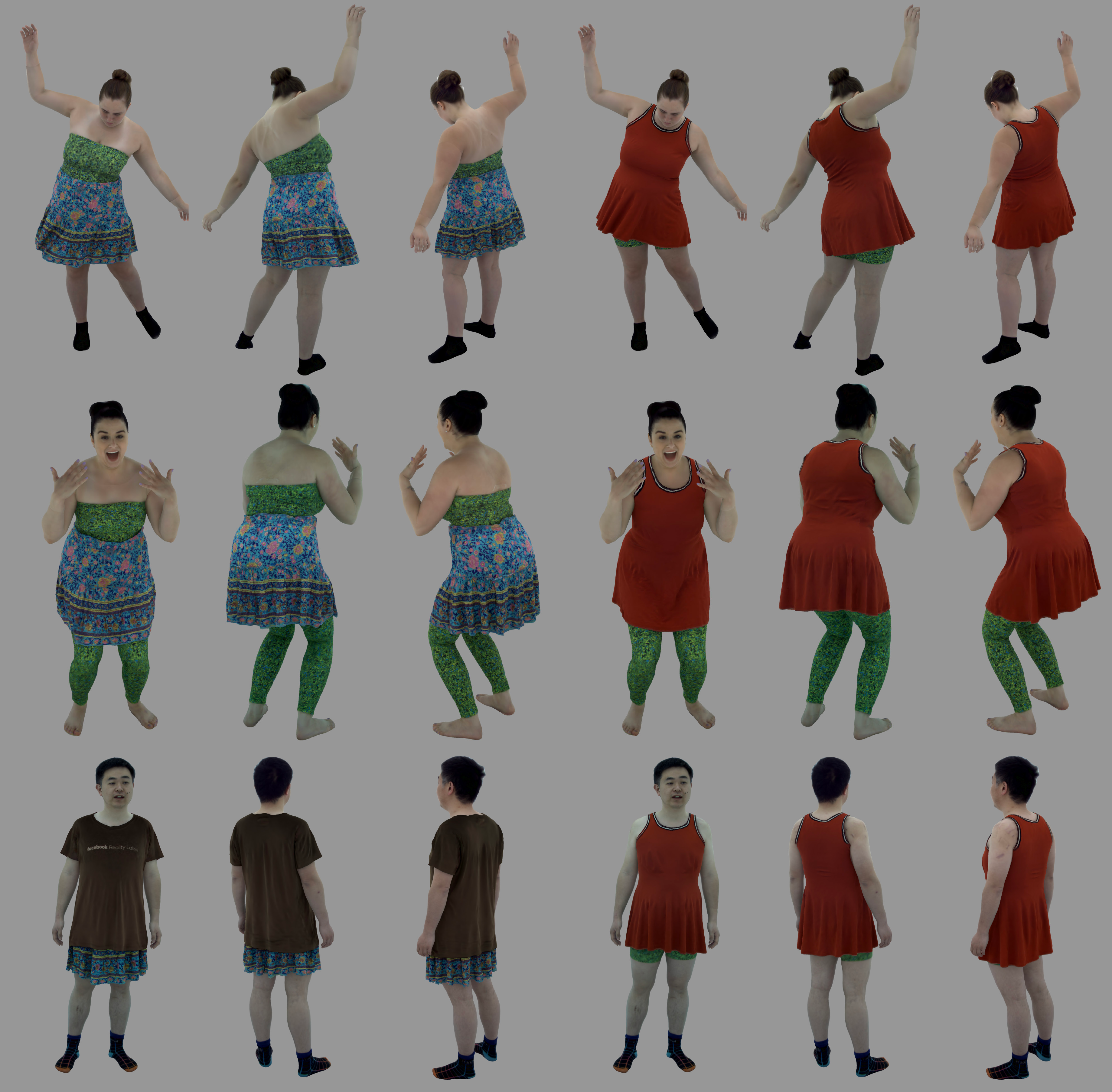}
    \caption{Dressing novel avatars. In the top row, a minimally clothed body avatar of the same actor as the original capture is shown in a skirt and dress. The middle and bottom rows show dressed avatars created from other actors. The bottom left results are dressed with two pieces of clothing together, a skirt and T-shirt.}
    \label{fig:dressing}
\end{figure*}

\begin{figure*}[t]
    \centering
    \includegraphics[width=\linewidth]{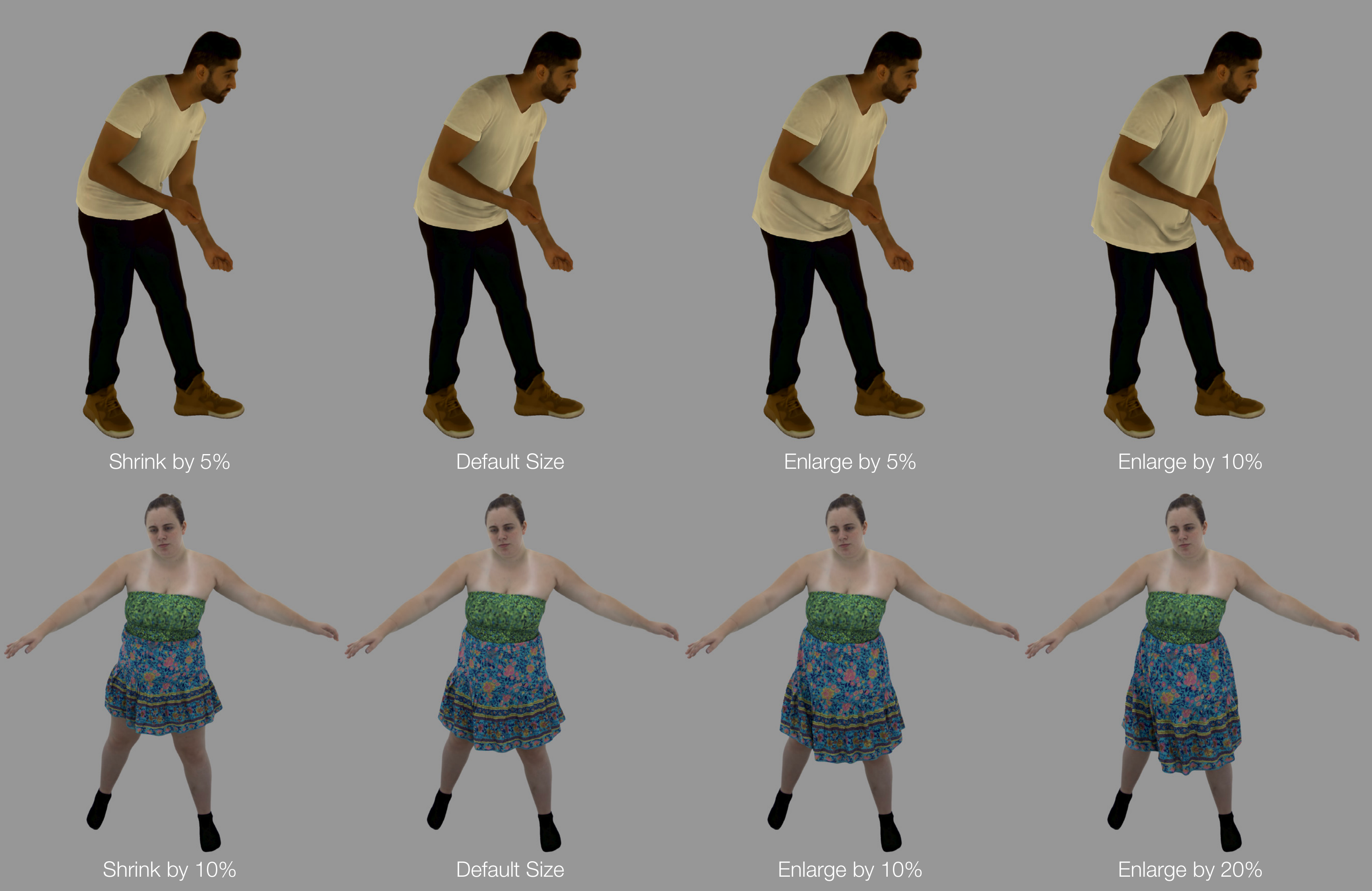}
    \caption{We edit garment sizes by scaling the template shape used in the physics-based simulation. Although the clothing appearance models are trained only using captured data of a particular size, they generalize well to the edited garments of different sizes.}
    \label{fig:edit-size}
\end{figure*}

\subsection{Pose-Driven Animation}
\label{sec:exp_animation}

In this section, we present pose-driven animation results for the four garments on top of the original body avatars that are captured together with the clothing. These results are generated by the complete animation pipeline shown in Fig.~\ref{fig:method_overview}. We make modest efforts to select the physical parameters for the simulation to roughly resemble the material behavior in the training sequence.

We compare the results with the output of the full method of Clothing Codec Avatars \cite{xiang2021modeling}. Some examples are shown in Fig.~\ref{fig:animation} and full animation results are shown in the supplementary video. The most obvious advantage of our method is the temporal dynamics, especially for the loose clothing that does not exactly follow the body motion. Although clothing is modeled explicitly in a separate layer in \cite{xiang2021modeling}, the mapping from body motion to clothing dynamics is a highly complex, temporally dependent function that is not easy to learn well. Our approach further separates the problem into dynamics and appearance to best capture both these aspects.

\subsection{Application: Dressing Avatars}
\label{sec:exp_dressing}

Our formulation naturally enables garment animations on top of base body avatars that are different from the original captured sequence. Because we model the clothing separately from the body layer, we simply replace the base body avatar and apply the physics-based simulation (with scaling if necessary) and clothing appearance model in exactly the same way as the default pipeline. We show dressed avatars of the same identity (wearing a tight capture suit) and different identities. The animation results are shown in Fig. \ref{fig:dressing} and in the supplementary video.

\subsubsection{Application: Editing garment size} We edit the garment size by changing the rest length of the garment template in the cloth simulator and keeping the same set of physical parameters. We maintain the same mesh topology, so that our appearance model can be directly applied to the edited garment. The animation results after size editing are shown in Fig. \ref{fig:edit-size}. This manipulation is only possible because both the physics-based simulator and our appearance model generalize well to unseen clothing configurations.

\subsection{Runtime Analysis}
\label{sec:exp_runtime}

In this section we report the runtime for the key modules of our algorithm: the physics-based cloth simulator, the base body avatar and the deep clothing appearance model.

We use a GPU-based cloth simulator with the eXtended Position Based Dynamics (XPBD) formulation~\cite{macklin2016xpbd}. When simulating a garment of 35k vertices on top of a body sequence of 9k vertices, the average runtime per frame for the simulation solver is between 8~ms $\sim$ 10~ms when taking 20 steps per frame on a Nvidia Tesla V100 GPU.

The base body avatar and the clothing appearance models are implemented using the PyTorch deep learning framework \cite{paszke2019pytorch}. For deployment, we use PyTorch Just-In-Time (JIT) compilation to improve performance. The average runtime for those two modules combined together is 78~ms on a Nvidia RTX 3090 GPU on a Lambda workstation. We use multiple GPUs to parallelize the inference. For example, we can achieve $>30$ fps with 3 GPUs in parallel. In the supplementary video, we show a demo of viewing an animated sequence in a VR headset, where the cloth simulation is precomputed and the body avatars and clothing appearance are rendered in real-time. Although we currently run the cloth simulator and the neural networks separately, given the modest computation for each of the three modules, our pipeline has the potential to be integrated into a complete real-time 
online system. 

\section{Discussion}

In this work, we present efficient full-body clothed avatars with physically plausible dynamics and photorealistic appearance. Our method achieves high-fidelity registration of dynamic cloth geometry and learns a deep clothing appearance model based on the registered geometry for training. At test time, the clothing is deformed through physics-based simulation and is rendered by the appearance model. To bridge the gap between tracked cloth geometry at training time and simulated clothing geometry at test time, we design an input representation and an architecture that consists of three modules inspired by the rendering equation. Once a garment is modeled, our system can use it to \textit{dress} a novel avatar. 

\subsubsection*{Limitations.} Clothing has a very large space of geometric and appearance variations. Although our method handles single-layer loose clothing much better than prior work, it may have difficulty in dealing with large folds of extremely loose clothing such as a Kimono or multi-layered clothing such as a coat and dress where the inner layers are significantly occluded. The photometric registration may fail when the feature matching is inaccurate due to bad initialization and/or repeated patterns. Some possibilities for improvement include using simulated clothing data with high-quality offline rendering to train the appearance model, or to utilize clothing with printed markers for registration~\cite{white2007capturing,halimi2022garment}.

In this work we manually adjust the physical parameters of the simulated garments. Optimization-based parameter estimation may improve the accuracy by leveraging recent advancements in differentiable cloth simulators \cite{liang2019differentiable,li2022diffcloth}. In addition, the physics-based cloth simulation may face challenges in handling highly complicated hand-cloth interactions such as dragging and pinching, due to the real-time constraint imposed on the simulation.

Although the clothing appearance model is trained using the data captured for only one subject, the shadow branch shows reasonable generalization ability when used to dress novel avatars because it is influenced by the body shape only indirectly through the occlusion of rays. Nevertheless, how the model works for extreme body shapes remains untested due to the limited availability of minimally clothed data. A systematic investigation of shadowing across different garments and identities is an interesting future direction. Furthermore, we do not consider complicated interreflection of lighting between clothing and body and between the folds of clothing. We also assume consistent illumination conditions when dressing novel avatars. When this assumption does not hold, slight inconsistency can be observed, for example, in the two-cloth animation results (T-shirt and skirt). In the future, we would like to relax this constraint by incorporating a relighting capability~\cite{sai2021relighting} to our avatar and clothing models with an updated capture setup.

\begin{acks}
We thank Anuj Pahuja for his contribution in exporting the clothing appearance model and building the VR demo. Donglai Xiang is supported by the Meta AI Mentorship (AIM) program.
\end{acks}

\bibliographystyle{ACM-Reference-Format}
\bibliography{sample-bibliography}

\end{document}